\documentclass[11pt, preprint]{article}   

\usepackage{amsmath}    
\usepackage{amsfonts}  
\usepackage{amssymb}
\usepackage{graphicx}   
\usepackage[T1]{fontenc}
\usepackage[utf8x]{inputenc}

\usepackage{float}   

\usepackage{color}   
\usepackage{soul}    
\usepackage{ulem}    
\usepackage{lipsum} 
\usepackage[a4paper]{geometry} 
\definecolor{lightblue}{rgb}{0.8,0.9,1} 

\def\signed #1{{\leavevmode\unskip\nobreak\hfil\penalty50\hskip2em
  \hbox{}\nobreak\hfil#1%
  \parfillskip=0pt \finalhyphendemerits=0 \endgraf}}

\newsavebox\mybox
\newenvironment{aquote}[1]
  {\savebox\mybox{#1}\begin{quote}\it}
  {\signed{\usebox\mybox}\end{quote}}

\usepackage[colorinlistoftodos]{todonotes}

\begin{document}

\title{Replacing the singlet spinor of the EPR-B experiment in the configuration space with two single-particle spinors in physical space}

\author{Michel Gondran$^1$ and Alexandre Gondran$^2$\\\small
$^1$ University Paris Dauphine, Lamsade, 75 016 Paris, France\\\small
\texttt{michel.gondran@polytechnique.org}\\\small
$^2$ \'Ecole Nationale de l'Aviation Civile, 31000 Toulouse, France\\\small
\texttt{alexandre.gondran@enac.fr}}

\author{\begin{tabular}{cc}
         Michel Gondran & Alexandre Gondran\\
         \small University Paris Dauphine, Lamsade, Paris, France & \small\'Ecole Nationale de l'Aviation Civile, Toulouse, France\\
         \small\texttt{michel.gondran@polytechnique.org} & \small\texttt{alexandre.gondran@enac.fr}
        \end{tabular}
}

\date{}
\maketitle

\begin{abstract}
 Recently, for spinless non-relativistic particles, Norsen, Marian and Oriols~\cite{Norsen2010,Norsen2014} show that in the de Broglie-Bohm interpretation it is possible to replace the wave function in the configuration space by single-particle wave functions in physical space. In this paper, we show that this replacment of the wave function in the configuration space by single-particle functions in the 3D-space is also possible for particles with spin, in particular for the particles of the EPR-B experiment, the Bohm version of the Einstein-Podolsky-Rosen experiment. 
\end{abstract}

\section{Introduction}

A major difficulty of the wave function interpretation of N particles in quantum mechanics is its definition in a 3N-dimensional configuration space. Since the Solvay Conference in 1927, de Broglie and Schr\"{o}dinger considered the wave function of N particles introduced by Schr\"{o}dinger in the 3N-dimensional configuration space as fictitious and proposed  to replace it by  N single-particle wave functions in 3D-space:

\begin{aquote}{Louis de Broglie~\cite{deBroglie1927}, cited by Norsen~\cite{Norsen2010}}
 "It appears to us certain that if one wants to \textit{physically} represent the evolution of a system of N corpuscles, one must consider the propagation of N waves in space, each  N propagation being determined by the action of the N-1 corpuscles connected to the other waves.  Nevertheless, if one focusses one's attention only on the corpuscles, one can represent their states by a point in configuration space, and one can try to relate the motion of this representative point to the propagation of a fictitious wave $\Psi $ in configuration space. It appears to us very probable that the wave
\begin{equation*}
\Psi=a(q_1, q_2,...,q_n) cos\frac{2\pi}{h}\varphi(t, q_1,...q_n),
\end{equation*}	
"a solution of the Schr\"{o}dinger equation, is only a fictitious wave, which in the \textit{Newtonian approximation}, plays for the representative point of the system in configuration space the same role of pilot wave and of probability wave that the wave $ \Psi$ plays in ordinary space in the case of a single material point."
\end{aquote}


\begin{aquote}{Erwin Schr\"{o}dinger~\cite{Schrodinger1927}, cited by Norsen et al.~\cite{Norsen2014} p.26}
 "This use of the q-space [configuration space] is to be seen only as a mathematical tool, as it is often applied also in the old mechanics; ultimately... the process to be described is one in space and time."
\end{aquote}

However, this program to replace the wave function in a 3N-dimensional configuration space by N single-particle wave functions was prematurely abandoned. It was recently re-opened by Norsen, Marian and Oriols \cite{Norsen2010, Norsen2014}. For spinless non-relativistic particles, these authors show that it is possible in the de Broglie-Bohm pilot-wave theory to replace the wave function in the configuration space by N single-particle wave functions in physical space \cite{Norsen2014}. These N wave functions in 3D space are the N \textit{conditional wave functions} of a subsystem introduced by D\"{u}rr, Goldstein and Zanghi \cite{Durr1992, Durr2004}.  For a N-particle wave function $\Psi(x_1, x_2,...,x_N,t) $, the N conditional wave functions are:
\begin{equation*}
\Psi_1(x,t)= \Psi(x, x_2,...,x_N,t)\vert_{x_2=X_2(t); x_N=X_N(t)}
\end{equation*}	
\begin{equation*}
\Psi_2(x,t)= \Psi(x_1, x,...,x_N,t)\vert_{x_1=X_1(t); x_N=X_N(t)}
\end{equation*}	
\begin{equation*}
\Psi_N(x,t)= \Psi(x_1,...,x_{N-1},x,t)\vert_{x_1=X_1(t); x_{N-1}=X_{N-1}(t)}
\end{equation*}	
where  $X_i(t)$ is the position of the particle $i$ at time $t$ in the Bohmian mechanics. The evolutions of these positions $X(t)= \lbrace X_1(t), X_2(t),..., X_N(t) \rbrace$ are given by the guidance formula:
\begin{equation*}
\dfrac{dX_i(t)}{dt}=\frac{\hbar}{m_i}Im \frac{\nabla_i\Psi}{\Psi}\vert_{\textbf{x}=\textbf{X}(t)}\equiv \frac{\hbar}{m_i}Im \frac{\nabla\Psi_i}{\Psi_i}\vert_{x=X_i(t)}
\end{equation*}

The aim of this paper is to show that this replacement of the wave function in the configuration space by single-particle functions in the 3D space is also possible in the de Broglie-Bohm interpretation for particles with spin, in particular for the particles in the singlet state of the EPR-B experiment, the Bohm version of the Einstein-Podolsky-Rosen experiment.  

To realize, in Bohmian mechanics, this decomposition, we use an explicit solution of the wave function of the EPR-B experiment. The first analytic expression of the wave function and of the probability density for the EPR-B experiment was done in 1987 by Dewdney, Holland and Kyprianidis~\cite{Dewdney1987b}  via
a complete integration of the two-body Pauli equation \textit{over
time and space}.
They give also the first causal interpretation of the EPR-B experiment~\cite{Dewdney1987b, Dewdney1986}. However, this interpretation presents a flaw: the spin module of each particle varied during the experiment from 0 to $\dfrac{\hbar}{2}$. The contribution  of this paper is, first, to correct this flaw by considering a spin module always equal to $\dfrac{\hbar}{2}$, and, second, to replace the singlet spinor of two entangled particles by two single-particle spinors.

The rest of the paper is organized as follows: section 2 recalls how Bohmian mechanics defines the spin of a quantum particle and how it interprets its measurement in a Stern-Gerlach apparatus.
The explicit solution of the two-body Pauli equation \textit{over
time and space} for EPR-B experiment, is presented in section 3. 
A new causal interpretation of the EPR-B experiment is proposed in section 4, correcting the flaw of the previous studies and allowing to replace the singlet spinor by two single-particle spinors.

\section{Spin and its measurement in Bohmian mechanics}

In the Copenhagen interpretation, the state of a spin 1/2 particle is given by the wave function $\Psi(\textbf{x},t)$, called Pauli spinor, which has two complex components $\Psi^{+}(\textbf{x},t) $ and $\Psi^{-}(\textbf{x},t) $. The non-relativist evolution of the spinor  $\Psi(\textbf{x},t)=\binom{\Psi^{+}(\textbf{x},t)}
                            {\Psi^{-}(\textbf{x},t)}$, for a neutral spin-1/2 particle with a mass $m$ and a magnetic moment $\mu$ in a magnetic field $\textbf{B}$, is given by the Pauli equation:
\begin{equation}\label{eq:Pauli}
    i\hbar \left( \begin{array}{c} \frac{\partial \Psi^{+}(\textbf{x},t)}{\partial t}
                                   \\
                                   \frac{\partial \Psi^{-}(\textbf{x},t)}{\partial t}
                  \end{array}
           \right)
    =-\frac{\hbar ^{2}}{2m}\left( \boldsymbol\nabla
                           \right)^{2}
                           \left( \begin{array}{c} \Psi^{+}(\textbf{x},t)
                                                   \\
                                                    \Psi^{-}(\textbf{x},t)
                                  \end{array}
                           \right)
     +\mu \textbf{B}\boldsymbol\sigma \left( \begin{array}{c} \Psi^{+}(\textbf{x},t)
                                                                  \\
                                                                  \Psi^{-}(\textbf{x},t)
                                                 \end{array}
                                          \right)
\end{equation}
where  $\boldsymbol\sigma=(\sigma_{1},\sigma_{2},\sigma_{3})$ corresponds to the three Pauli matrices.

In the de Broglie-Bohm interpretation, the wave function does not completely represent the state of the quantum particle and it is necessary to add the particle position $X(t)$. The evolution of the spinor is still given by the Pauli equation (\ref{eq:Pauli}) and the evolution of the position is given by the guidance formula introduced by Takabayasi~\cite{Takabayasi1954}, Bohm et al.~\cite{Bohm1955}:
 
\begin{equation}\label{eq:vitesse}
\dfrac{dX(t)}{dt}=\frac{\hbar}{2m \rho} Im{(\Psi^\dag\boldsymbol\nabla \Psi)}
\end{equation}	
where $\Psi^\dag=(\Psi^{+*}, \Psi^{-*})$ and $\rho=\Psi^\dag\Psi$.
A two-component spinor can be linked to the three Euler angles ($\theta$,$\varphi$,$\chi$) and we can write (cf. Bohm and Hiley~\cite{Bohm1993} p.206):
\begin{equation*}
\Psi(\textbf{x},t)=\left( \begin{array}{c} \Psi^{+}(\textbf{x},t)\\ \Psi^{-}(\textbf{x},t)
                  \end{array}
                  \right)
    =\sqrt{\rho} e^{i\frac{\chi}{2}} \left( \begin{array}{c} cos\frac{\theta}{2} e^{i\frac{\varphi}{2}}
                                   \\
                                  i sin\frac{\theta}{2} e^{-i\frac{\varphi}{2}}
                  \end{array}\right),
\end{equation*}	
where $\rho$, $\chi$, $\theta$ and $\varphi$ are functions of $\textbf{x}$ and t.
Bohm and al.\cite{Bohm1955} define the spin vector as
\begin{eqnarray}\label{eq:spinvector}
\textbf{s}(\textbf{x},t)=\frac{\hbar}{2\rho}\Psi^\dag(\textbf{x},t)\boldsymbol\sigma\Psi(\textbf{x},t)
=\frac{\hbar}{2}(sin\theta~ sin\varphi, sin\theta ~cos\varphi, cos\theta).
\end{eqnarray}
More properly, $\textbf{s}(\textbf{x},t)$ is a spin vector field: in each point of the 3D space, a vector is defined by the orientation of $\theta$ and $\varphi$. The spin vector of an individual particle defined by $\Psi(\textbf{x},0)$ and $X(0)$ is given by the Eq.~(\ref{eq:spinvector}) evaluated along its trajectory $X(t)$:
\begin{equation}\label{eq:spinvectorind}
 \textbf{s}(t)= \textbf{s }(\textbf{x},t)\vert_{\textbf{x}=X(t)}=\textbf{s}(X(t),t).
\end{equation}

The spin vector therefore depends on the spinor and on the position of the particle.
As D\"{u}rr, Goldstein and Zanghi remark~\cite{Durr2004}: \textit{"Unlike position,
spin is not primitive, i.e., no actual discrete degrees of freedom, analogous
to the actual positions of the particles, are added to the state
description in order to deal with "particles with spin". Roughly speaking,
spin is merely in the wave function"}.

In the de Broglie-Bohm interpretation, the initial wave function of a quantum particle, prepared with a spin vector having ($\theta_0$ ,$\varphi_0$ ) as Euler angles, is described by a spinor as the following:
\begin{eqnarray}\label{eq:psi-0}
    \Psi_{0}(x,z) &=& (2\pi\sigma_{0}^{2})^{-\frac{1}{2}}
                      e^{-\frac{x^2 +z^2}{4\sigma_0^2}}
                      \left( \begin{array}{c}\cos \frac{\theta_0}{2}e^{ i\frac{\varphi_0}{2}}
                                   \\
                                  \sin\frac{\theta_0}{2}e^{-i\frac{\varphi_0}{2}}
                  \end{array}
           \right)\\\label{eq:psi-0_bis}
           &=&(2\pi\sigma_{0}^{2})^{-\frac{1}{2}}
                      e^{-\frac{x^2 +z^2}{4\sigma_0^2}}\left(\cos \frac{\theta_0}{2}e^{ i\frac{\varphi_0}{2}}|+\rangle+\sin\frac{\theta_0}{2}e^{- i\frac{\varphi_0}{2}}|-\rangle\right)
\end{eqnarray}
corresponding to a pure state.

Quantum mechanics textbooks \cite{Feynman1965, CohenTannoudji1977, Sakurai1985,
LeBellac2006} do not take into account the spatial extension of the
spinor (\ref{eq:psi-0}) and simply use the simplified spinor
without spatial extension:
\begin{equation}\label{eq:psi-s}
    \Psi_{0} = \left( \begin{array}{c}\cos \frac{\theta_0}{2}e^{ i\frac{\varphi_0}{2}}
                                   \\
                                  \sin\frac{\theta_0}{2}e^{- i\frac{\varphi_0}{2}}
                  \end{array}
           \right).
\end{equation}
This spatial extension enables, in following the precursory works of Takabayasi \cite{Takabayasi1954}, Bohm et al. \cite{Bohm1955} and Dewdney and al. \cite{Dewdney1987b}, to taking into account the spin evolution during the measurement.
 Indeed, the difference in the evolution of the spatial
extension between the two spinor components has a key role
in the explanation of the measurement process with Bohmian mechanics (cf.~equation~(\ref{eq:fonctionapreschampmagnetique}) in Appendix).

In the Stern-Gerlach experiment, the spin of equations (\ref{eq:psi-0}) or (\ref{eq:psi-s}) is not directly measured; its measurement is obtained, after passing through an electromagnet during a time $\Delta t $, from the impact of the particle on a screen located $20~cm$ after the Stern-Gerlach electromagnet. This distance corresponds to the time required to separate the initial wave packet into two disjoint packets; this is the \textit{decoherence time} $t_D$ (cf.~equation (\ref{eq:tempsdecoherence}) in Appendix).

The measurement of spin (up or down along the z-axis) has no a pre-existing value before measurement. For the spinor (\ref{eq:psi-0}), the initial spin vector $\textbf{s}(X(0),0)=\frac{\hbar}{2}(\sin\theta_0\sin\varphi_0,\ \sin\theta_0 \cos\varphi_0,\ \cos\theta_0)$ does not depend on the initial position $X(0)= (x_0, y_0, z_0) $, but will evolve as $ \textbf{s}(X(t),t)$ differently during measurement depending on the initial position $z_0$ of the particle.

\begin{figure}[H]
\begin{center}
\includegraphics[width=0.7\linewidth]{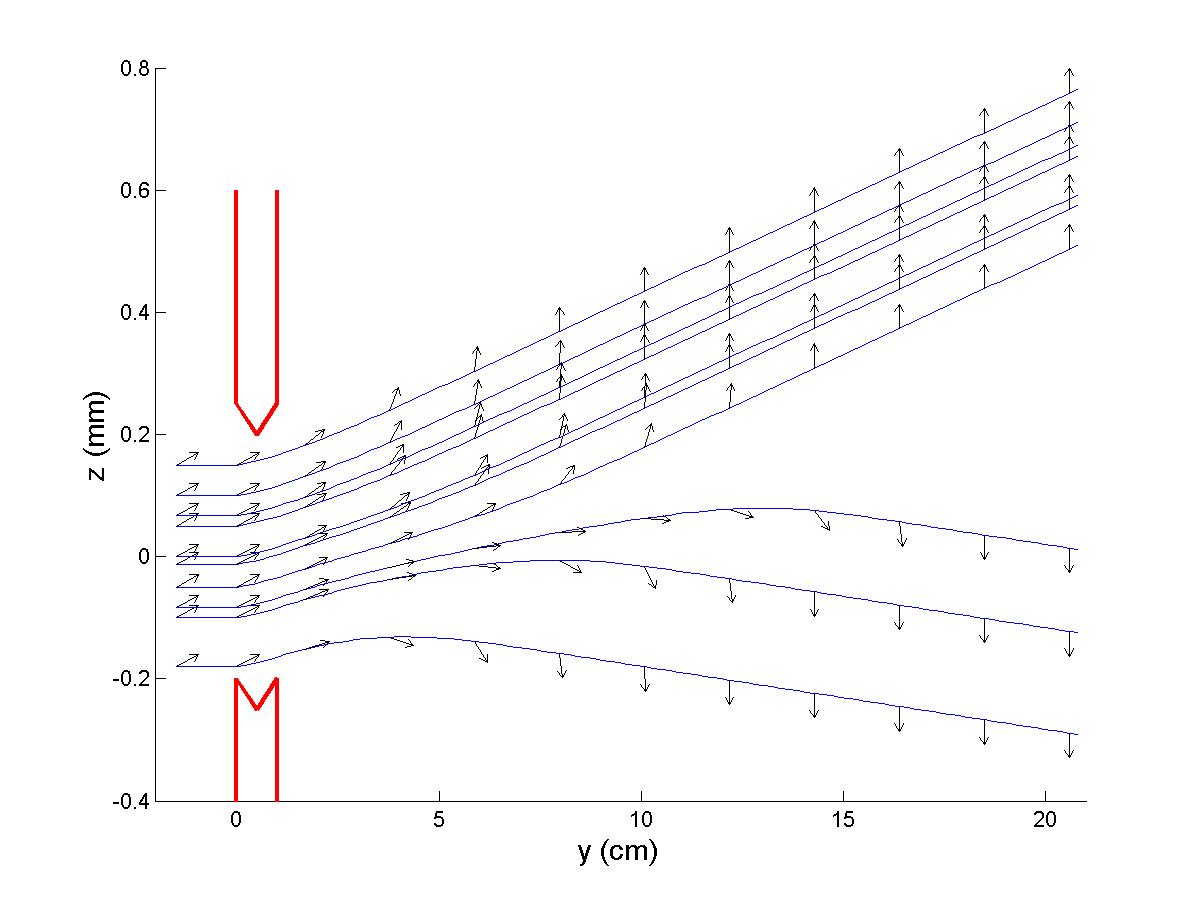}
\caption{\label{fig:SetG-10traj} Ten silver atom trajectories with
the same initial spinor orientation $(\theta_0=\frac{\pi}{3})$ and with 10 different initial positions $z_0$; arrows represent the spin orientation $\theta(z(t),t)$ along the trajectories.}
\end{center}
\end{figure}

Figure~\ref{fig:SetG-10traj} presents in the $(Oyz)$ plane, 10 trajectories of silver atoms having the same initial spinor orientations $(\theta_0=\frac{\pi}{3},\varphi_0=0)$ but having 10 different initial positions $z_0$. Those initial
positions $z_0$ have been randomly chosen from a Gaussian
distribution with standard deviation $\sigma_{0}$. The spin
orientation $\theta(z(t),t)$ of each atom is represented by arrows.
The final orientation, obtained after the decoherence time
$t_{D}$ (equation (\ref{eq:tempsdecoherence}) in Appendix), depends on the specific initial particle position $z_{0}$ in the
spinor with a spatial extension and on the initial angle
$\theta_{0}$ of the spin with the $z$-axis. In previous works~\cite{Dewdney1986,Gondran2005b}, 
$\theta(t_D)=+\frac{\pi}{2}$  is obtained if $z_0
> z^{\theta_{0}}$ and $\theta(t_D)=-\frac{\pi}{2}$ if $z_0
< z^{\theta_{0}}$ with
\begin{equation}\label{eq:seuilpolarization}
z^{\theta_{0}}=\sigma_0 \Phi^{-1}\left(\sin^{2}\frac{\theta_{0}}{2}\right)
\end{equation}
where $\Phi$ is the cumulative distribution of the standard normal distribution. If we ignore the position of the atom in its wave function,
we lose the determinism given by equation
(\ref{eq:seuilpolarization}).

In the de Broglie-Bohm interpretation, the "measured" value
is not a preexisting value. It is contextual and conforms to the Kochen and Specker
theorem~\cite{Kochen1967}.

Finally, the Bohmian mechanics proposes a clear interpretation of the spin 
measurement in quantum mechanics. There is interaction with the
measuring apparatus as Bohr said; and there is indeed a
minimum time required to measure. However this measurement and
this time do not have the meaning that is usually attributed to them.
The result of the Stern-Gerlach experiment is not the measure of
the spin projection along the $z$-axis, but the continuous orientation of the
spin either in the direction of the magnetic field gradient, or in
the opposite direction. It depends on the position of the particle
in the wave function. We have therefore a simple explanation for
the non-compatibility of spin measurements along different axes.
The measurement duration ($t\geqslant t_D$) is then the time necessary for the
particle to point its spin in the final direction.

\section{Explicit solution of the spinor in configuration space for the EPR-B experiment}

\begin{figure}
\begin{center}
\includegraphics[width=0.9\linewidth]{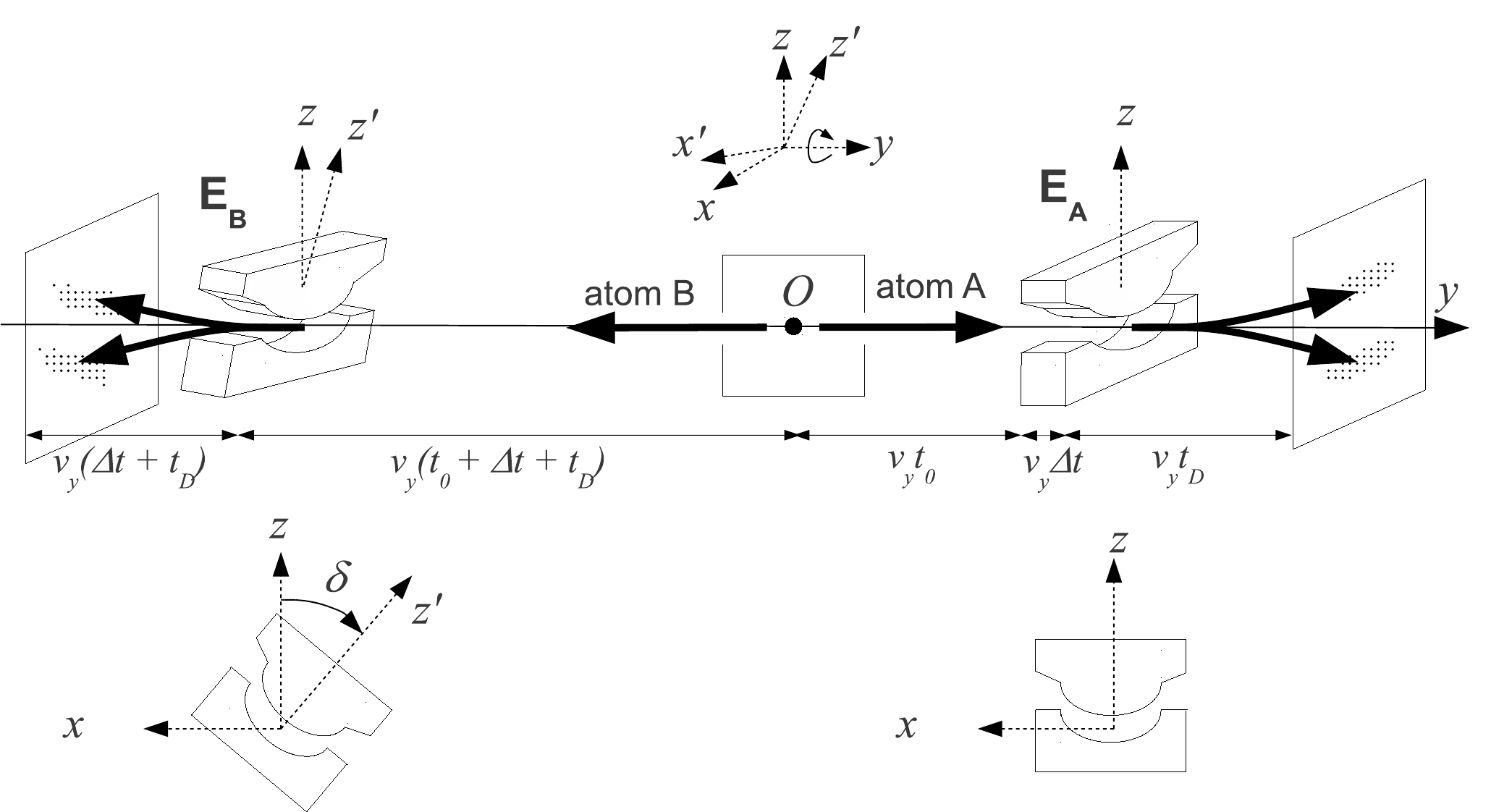}
\end{center}
\caption{\label{fig:expEPR}Schematic configuration of the EPR-B
experiment.}
\end{figure}

Figure \ref{fig:expEPR} presents the Einstein-Podolsky-Rosen-Bohm
experiment. A source $S$ creates in $O$ an entangled pair of identical atoms A
and B, with opposite spins. The atoms A and B split following
the $y$-axis in opposite directions (B with velocity $v_0$, A with velocity $-v_0$), and head towards two identical
Stern-Gerlach apparatuses $\mathcal{A}$ and $\mathcal{B}$.

The electromagnet $\mathcal{A}$ "measures" the A spin in the
direction of the $z$-axis and the electromagnet $\mathcal{B}$
"measures" the B spin in the direction of the $z'$-axis, which is
obtained after a rotation of an angle $\delta$ around the $y$-axis.

In most papers on EPR-B experiment, the initial wave function of the quantum system  composed of two entangled particles is the singlet spinor:
\begin{equation}\label{eq:7psi2}
    \Psi_{0} =\frac{1}{\sqrt{2}}(|+_{A}\rangle |-_{B}\rangle - |-_{A}\rangle | +_{B}\rangle)
\end{equation}
where $|\pm_{A}\rangle$ (resp. $|\pm_{B}\rangle$) are the eigenvectors
of the spin operators $S_{z_A}$ (resp. $S_{z_B}$) in
the $z$-direction pertaining to particle A (resp.B):  $S_{z_A}
|\pm_{A}\rangle= \pm (\frac{\hbar}{2})|\pm_{A}\rangle$
(resp. $S_{z_B} |\pm_{B}\rangle= \pm
( \frac{\hbar}{2})|\pm_{B}\rangle$).

More specifically, the initial singlet wave function has a spatial extension:
\begin{equation}\label{eq:7psi-1}
    \Psi_{0}(\textbf{r}_A,\textbf{r}_B) =\frac{1}{\sqrt{2}}f(\textbf{r}_A) f(\textbf{r}_B)(|+_{A}\rangle |-_{B}\rangle - |-_{A}\rangle | +_{B}\rangle)
\end{equation}
where $\textbf{r}= (x,z)$ and $f(\textbf{r})=(2\pi\sigma_{0}^{2})^{-\frac{1}{2}}
 e^{-\frac{x^2 + z^2}{4\sigma_0^2}}$.
This spatial extension is essential to solve correctly the Pauli equation in space. 
Moreover, in Bohmian mechanics, the spatial extension is necessary to take into account particle position.

In the Copenhaguen interpretation, the result of the simultaneous measurement of two spins is obtained directly from the initial wave function (\ref{eq:7psi2}) written in the basis of the eigenvectors of the measuring operators, using for this the quantum-mechanics  measurement postulates. For this, we make a coordinate change of the particle B, placing it in the plane $x'Oy'$ obtained from $Oz$ by rotation $\delta$ around $Oy$. So we have:
\begin{equation*}
x_B= x'_B \cos\delta+ z'_B \sin\delta~~~~\textnormal{and}~~~~z_B= - x'_B
\sin\delta+ z'_B \cos\delta.
\end{equation*}
 In the new basis $|\pm'_{B}\rangle$ of the eigenvectors of the operator
$\sigma_{z'_B}$, the spinors $|\pm_{B}\rangle$ are written:
\begin{equation}\label{eq:changebase}
|+_{B}\rangle =\cos\frac{\delta}{2}|+'_{B}\rangle +
\sin\frac{\delta}{2}|-'_{B}\rangle~~~~\textnormal{and}~~~~ |-_{B}\rangle =-
\sin\frac{\delta}{2}|+'_{B}\rangle +
\cos\frac{\delta}{2}|-'_{B}\rangle.
\end{equation}
The initial wave function (\ref{eq:7psi2}) is then written:
\begin{equation}\label{eq:7psirotation}
\Psi_{0}= \frac{1} {\sqrt{2}} (-\sin\frac{\delta}{2}|+_{A}\rangle| +'_{B}\rangle
+\cos\frac{\delta}{2} |+_{A}\rangle| -'_{B}\rangle
-\cos\frac{\delta}{2} |-_{A}\rangle| +'_{B}\rangle
-\sin\frac{\delta}{2}|-_{A}\rangle| -'_{B}\rangle).
\end{equation}

Since, in the EPR-B experiment, A spin is measured along the $z$-axis and that of B along the $z'$-axis, the wave function (\ref{eq:7psirotation}) predicts the following probabilities for the measuring torques:
\begin{equation}\label{eq:7probaEPR}
P(+,+)=P(-,-)=\frac{1}{2}\sin^2\frac{\delta}{2}~~,~~P(+,-)=P(-,+)=\frac{1}{2}\cos^2\frac{\delta}{2}
\end{equation}
where $P(+,-)$, for example, corresponds to the probability to find A with the spin up (+) and B with the spin down ($-$).

The same probabilities are obtained if we take the initial singlet (\ref{eq:7psi-1}) with a spatial extension. Indeed as $f(\textbf{r}_B)=f(\textbf{r}'_B)$, we have: 
\begin{eqnarray} \nonumber
\Psi_{0}(\textbf{r}_A,\textbf{r}'_B)=\left.\frac{1} {\sqrt{2}} f(\textbf{r}_A)
f(\textbf{r}'_B)\right(&-&\sin\frac{\delta}{2}|+_{A}\rangle| +'_{B}\rangle
+\cos\frac{\delta}{2} |+_{A}\rangle| -'_{B}\rangle\\
&-&\left.\cos\frac{\delta}{2} |-_{A}\rangle| +'_{B}\rangle
-\sin\frac{\delta}{2}|-_{A}\rangle| -'_{B}\rangle\right)\label{eq:7psirotationb}
\end{eqnarray}
and the calculation of $P(+,-)$ for example is made by integration~: 
$$\displaystyle P(+,-)=\int \frac{1}{2}\cos^2\frac{\delta}{2} |f(\textbf{r}_A)|^2
|f(\textbf{r}'_B)|^2 d\textbf{r}_A  d\textbf{r}'_B=
\frac{1}{2}\cos^2\frac{\delta}{2}.$$

In the de Broglie-Bohm interpretation, the postulates of quantum-mechanics measurement are not used and results of the measurement are obtained by calculating the evolution of the wave function in interaction with measuring apparatuses (see Appendix for numerical values).

Let us consider the wave function of the two particles A and B of the EPR-B experiment in the configuration space. These are two identical particles:both are electrically neutral, with magnetic moments $\mu$, and are respectively subject to magnetic fields
$\textbf{B}^{\mathcal{A}}$ and $\textbf{B}^{\mathcal{B}}$. This wave function $\Psi(\textbf{r}_A, \textbf{r}_B, t)$ admits 4 components
$\Psi^{a,b}(\textbf{r}_A, \textbf{r}_B, t)$ on the basis $[|\pm_{A}\rangle,|\pm_{B}\rangle] $ with $a=\pm$ and $b=\pm$, and its evolution is given by the
two-body Pauli equation (see Holland~\cite{Holland1993} p. 417 and D\"{u}rr et al.~\cite{Durr2004}). In Einstein notation this is expressed as:
\begin{eqnarray}
    i\hbar \frac{\partial \Psi^{a,b}}{\partial t}
    =\left(-\frac{\hbar^2}{2 m}\left( \boldsymbol\nabla_A
                           \right)^{2} -\frac{\hbar^2}{2 m}\left( \boldsymbol\nabla_B
                           \right)^{2}\right)\Psi^{a,b}
    +\mu B^{\mathcal{A}}_j (\textbf{r}_A)(\sigma_j)_{c}^{a}\Psi^{c,b}
    +\mu B^{\mathcal{B}}_j (\textbf{r}_B)(\sigma_j)_{d}^{b}\Psi^{a,d}\label{eq:7Paulideuxcorps1}
\end{eqnarray}
with j=1 to 3. The initial conditions are:
\begin{equation}\label{eq:7Paulideuxcorps2}
\Psi^{a,b}(\textbf{r}_A, \textbf{r}_B,
0)=\Psi_0^{a,b}(\textbf{r}_A, \textbf{r}_B)
\end{equation}
where the
$\Psi_0^{a,b}(\textbf{r}_A, \textbf{r}_B)$ correspond to the
singlet state (\ref{eq:7psi-1}).

One of the difficulties of the interpretation of the EPR-B
experiment is the existence of two simultaneous measurements. By
doing these measurements one after the other as proposed in 1987 by Dewdney, Holland and Kyprianidis~\cite{Dewdney1986}, 
it facilitates the interpretation and calculation of the experiment. 
That is the purpose of the two-step version of the EPR-B experiment studied below.
 The latter experiment is equivalent to the previous experiment and gives the usual correlations (\ref{eq:7probaEPR}) of the initial EPR-B experiment.

Classic treatments of the EPR-B experiment within Bohmian mechanics~\cite{Holland1993, Dewdney1987b, Holland1988b} focus only on final calculations in order to ensure consistency with experimental results. In our view, intermediate formulas (those after the first step) are also very interesting to present. That is why we detail below the complete calculations and the conclusions after each step.

\subsection{First step: Measurement of A spin}

In the first step we make, on a pair of entangled particles A and B, a Stern-Gerlach "measurement" for atom A, then in the second step a  Stern-Gerlach "measurement" for atom B.

Consider that at time $t_0$ the particle A arrives at the entrance
of electromagnet $\mathcal{A}$. $\triangle t$ is the
duration of crossing electromagnet $\mathcal{A}$ and $t$ is the time after
the $\mathcal{A}$ exit. 
At time $t_0+ \triangle t + t$, wave function (\ref{eq:7psi-1}) becomes~\cite{Gondran2009}:
\begin{eqnarray}
\Psi(\textbf{r}_A, \textbf{r}_B, t_0 + \triangle t+ t )&=& \frac{1}
{\sqrt{2}}f(\textbf{r}_B)\left(f^{+}(\textbf{r}_A,t) |+_{A}\rangle | -_{B}\rangle -
f^{-}(\textbf{r}_A,t) |-_{A}\rangle |+_{B}\rangle\right)\label{eq:7psiexperience1}
\end{eqnarray}
with
\begin{equation}\label{eq:7fonction}
f^{\pm}(\textbf{r},t)\simeq f(x, z \mp z_\triangle \mp ut)
e^{i(\frac{\pm muz}{\hbar}+ \varphi^\pm (t))}.
\end{equation}
where $z_{\Delta}$ and $u$ are defined in Appendix by equation (\ref{eq:zdeltavitesse}).

The atomic density $\rho(z_A, z_B,t_0 + \Delta t + t)$ is found by
integrating $\Psi^\dag(\textbf{r}_A,\textbf{r}_B,t_0+\triangle t+t) \times \Psi(\textbf{r}_A, \textbf{r}_B, t_0 + \triangle t+ t)$ on
$x_A$ and $x_B$:

\begin{eqnarray}
    \rho(z_A, z_B,t_0 + \Delta t+ t) &=& \left((2\pi\sigma_0^2)^{-\frac{1}{2}}
                  e^{-\frac{(z_B)^2}{2\sigma_0^2}}\right)
                  \label{eq:7densiteapreschampmagnetiqueAB}\\
     &&\times\left((2\pi\sigma_0^2)^{-\frac{1}{2}}
                  \frac{1}{2}\left(e^{-\frac{(z_A-z_{\Delta}- ut)^2}{2\sigma_0^2}}+
                  e^{-\frac{(z_A+z_{\Delta}+
                  ut)^2}{2\sigma_0^2}}\right)\right).\nonumber
\end{eqnarray}

Our previous studies~\cite{Gondran2009} on the EPR-B experiment are based on equations~(\ref{eq:7psiexperience1}) and (\ref{eq:7densiteapreschampmagnetiqueAB}), which are not explicitly presented in the classic Bohmian framework.
We deduce from (\ref{eq:7densiteapreschampmagnetiqueAB}) that the beam of particles A is divided into two parts, while
the B beam of particles is not, and remains stable. 
Moreover, we note that the space quantization of particle A is
identical to that of an  free particle in a single Stern-Gerlach
apparatus: the distance $\delta z= 2(z_{\Delta}+ ut)$ between the
two spots $N^+$ (spin +) and $N^-$ (spin $-$) of a set of
particles A is the same as the distance between the two spots $N^+$
and $N^-$ of a set of particles in a single Stern-Gerlach
experiment~\cite{Gondran2005b}, cf. (\ref{eq:densiteapreschampmagnetique}) in Appendix. 
We finally deduce from (\ref{eq:7densiteapreschampmagnetiqueAB})
that:
\begin{itemize}
    \item the density of B is not affected by the "measurement" of A,
	\item the density of A is the same, whether particle A is entangled
with B (\ref{eq:7densiteapreschampmagnetiqueAB}) or not.	
\end{itemize}
These two results can be tested experimentally.
We also conclude from (\ref{eq:7psiexperience1})
that the spins of A and B remain opposite throughout the experiment. 
This analysis of the first step was not made in the previous Bohmian approaches. It also provides the possibility of replacing the single spinor of two entangled particles with two independent single-particle spinors, plus an interaction-at-a-distance that maintains the two spin vectors in opposite directions.

\subsection{Second step: "Measurement" of B spin.}

After a first step of a Stern-Gerlach "measurement" on atom A, between $t_0$ and
$t_1=t_0+\triangle t+t_D$, the second step corresponds to a  Stern-Gerlach "measurement"
on atom B, with an electromagnet $\mathcal{B}$ forming an angle
$\delta$ with $\mathcal{A}$ between $t_1$ and
$t_1+ \triangle t + t_D$.

At time $t_1$, the wave function in configuration space is given by
(\ref{eq:7psiexperience1}) with $t=t_D$. In the new basis [$|\pm_{A}\rangle,|\pm'_{B}\rangle$], this wave function is written:
\begin{eqnarray}\label{eq:7psirotationB}
\left.\Psi(\textbf{r}_A, \textbf{r}'_B,t_1)= \frac{1}{\sqrt{2}}f(\textbf{r}'_B)\right[
&-&\sin\frac{\delta}{2}f^{+}(\textbf{r}_A,t_D)|+_{A}\rangle| +'_{B}\rangle
+\cos\frac{\delta}{2} f^{+}(\textbf{r}_A,t_D)|+_{A}\rangle| -'_{B}\rangle\nonumber\\
&-&\left.\cos\frac{\delta}{2} f^{-}(\textbf{r}_A,t_D)|-_{A}\rangle| +'_{B}\rangle
-\sin\frac{\delta}{2}f^{-}(\textbf{r}_A,t_D)|-_{A}\rangle| -'_{B}\rangle\right].
\end{eqnarray}

After the measurement of B at its exit of 
magnetic field $\mathcal{B}$, at time $t_1+ \triangle t + t_D$, the
wave function (\ref{eq:7psirotationB}) becomes:
\begin{eqnarray}\label{eq:7psirotationBB}
\left.\Psi(\textbf{r}_A, \textbf{r}'_B,t_1 + \triangle t +t_D)= \frac{1} {\sqrt{2}} \right[&-&\sin\frac{\delta}{2}f^{+}(\textbf{r}_A,t_D)f^{+}(\textbf{r}'_B,t_D)|+_{A}\rangle| +'_{B}\rangle\nonumber\\
&+&\cos\frac{\delta}{2} f^{+}(\textbf{r}_A,t_D)f^{-}(\textbf{r}'_B,t_D)|+_{A}\rangle| -'_{B}\rangle
\nonumber\\
&-&\cos\frac{\delta}{2} f^{-}(\textbf{r}_A,t_D)f^{+}(\textbf{r}'_B,t_D)|-_{A}\rangle| +'_{B}\rangle
\nonumber\\
&-&\left.\sin\frac{\delta}{2}f^{-}(\textbf{r}_A,t_D)f^{-}(\textbf{r}'_B,t_D)|-_{A}\rangle| -'_{B}\rangle\right].
\end{eqnarray}

Equation (\ref{eq:7psirotationBB}) predicts probabilities of (\ref{eq:7probaEPR}).
The calculation of $P(+,-)$, for example, is made by integration:
$\displaystyle P(+,-)=\int \frac{1}{2}\cos^2\frac{\delta}{2} |f^{+}(\textbf{r}_A, t_D)|^2
|f^{-}(\textbf{r}'_B, t_D)|^2 d\textbf{r}_A  d\textbf{r}'_B=
\frac{1}{2}\cos^2\frac{\delta}{2}$.

Similarly to Holland, we obtain, for spatial quantization and
correlations of spins in this two-step version of the EPR-B
experiment, the same results as in the EPR-B experiment. The EPR correlations are therefore obtained only by solving the two-body Pauli equation  without using the quantum-measurement postulates.

\section{Two single-particle spinors in the EPR-B experiment}

In this section, we present the principal contribution of the paper: how to replace the singlet spinor of EPR-B experiment with two single-particle spinors plus an interaction-at-a-distance that maintains the two spin vectors in opposite directions.
When the entangled pair of particles A and B is created, we assume that each particle has the initial wave function:
$\Psi_0^A(\textbf{r}_A, \theta^A_0,
\varphi^A_0)$ and $\Psi_0^B(\textbf{r}_B, \theta^B_0,
\varphi^B_0)$ like in equation (\ref{eq:psi-0_bis}): 
\begin{equation}\label{eq:A}
\Psi_0^A(\textbf{r}_A, \theta^A_0, \varphi^A_0)= f(\textbf{r}_A)
\left(\cos\frac{\theta^A_0}{2}|+_{A}\rangle +
\sin\frac{\theta^A_0}{2}e^{i \varphi^A_0}|-_{A}\rangle\right)    
\end{equation}
and
\begin{equation}\label{eq:B}
\Psi_0^B(\textbf{r}_B , \theta^B_0, \varphi^B_0)= f(\textbf{r}_B)
\left(\cos\frac{\theta^B_0}{2}|+_{B}\rangle +
\sin\frac{\theta^B_0}{2}e^{i \varphi^B_0}|-_{B}\rangle\right).   
\end{equation}
Moreover those spinors have opposite spins: 
$\theta_0^B= \pi-\theta_0^A$, $\varphi_0^B= \varphi_0^A -\pi$. We treat the
dependence on $y$ classically: speed $- v_0$ for A and $ v_0$ for B.
Then the Pauli principle tells us that the two-body wave function
must be antisymmetric; it is written: 
\begin{equation}\nonumber
 \Psi_0(\textbf{r}_A,\theta_A, \varphi_A,\textbf{r}_B,\theta_B, \varphi_B)=
 \Psi^0_A(\textbf{r}_A,\theta_A, \varphi_A)\Psi^0_B(\textbf{r}_B,\theta_B,
  \varphi_B)-\Psi^0_A(\textbf{r}_B,\theta_B, \varphi_B)\Psi^0_B(\textbf{r}_A,\theta_A, \varphi_A)
\end{equation}
i.e.
$\Psi_0(\textbf{r}_A,\theta_A, \varphi_A, \textbf{r}_B,\theta_B,
\varphi_B)=
f(\textbf{r}_A)f(\textbf{r}_B)[(\cos\frac{\theta_A}{2}|+_{A}\rangle
+ \sin\frac{\theta_A}{2}e^{i
\varphi_A}|-_{A}\rangle)(\cos\frac{\theta_B}{2}|+_{B}\rangle +
\sin\frac{\theta_B}{2}e^{i \varphi_B}|-_{B}\rangle)-
(\cos\frac{\theta_B}{2}|+_{A}\rangle + \sin\frac{\theta_B}{2}e^{i
\varphi_B}|-_{A}\rangle)(\cos\frac{\theta_A}{2}|+_{B}\rangle +
\sin\frac{\theta_A}{2}e^{i \varphi_A}|-_{B}\rangle)]$, 
and after calculation we obtain the same singlet state as (\ref{eq:7psi-1}), factor-wise:
\begin{equation}\nonumber
 \Psi_0(\textbf{r}_A,\theta_A, \varphi_A,\textbf{r}_B,\theta_B, \varphi_B)= - e^{i \varphi_A} f(\textbf{r}_A)f(\textbf{r}_B)(|+_{A}\rangle
|-_{B}\rangle - |-_{A}\rangle|+_{B}\rangle)
\end{equation}


The assumption of the existence of initial wave functions $\Psi_0^A(\textbf{r}_A, \theta^A_0,
\varphi^A_0)$ and $\Psi_0^B(\textbf{r}_B, \theta^B_0,
\varphi^B_0)$ (equations~(\ref{eq:A}-\ref{eq:B})) is consistent with singlet state (\ref{eq:7psi-1}) and new in Bohmian interpretations.
It is important to note that each entangled pair of atoms has a different value of $(\theta^A_0,\varphi^A_0)$ and thus ($\theta_0^B= \pi-\theta_0^A$, $\varphi_0^B= \varphi_0^A -\pi$). At each emission of one EPR-B pair, the initial spin directions are \textit{unknown}: $\theta_0^A$ have a uniform distribution over [0,$\pi$] and $\varphi_0^A$ have a uniform distribution over [0,$2\pi$], according to the invariance in all rotations in 3D space.



This interpretation differs from the classic Bohmian interpretation.
Indeed, in Dewdney and al.~\cite{Dewdney1987b, Holland1988b}, Bohm and Hiley~\cite{Bohm1993} (p. 226) and Holland~\cite{Holland1993} (p. 417 and 467), the spin vectors are defined  by a generalization of the definition (\ref{eq:spinvector}) applied to the singlet wave function, and no to a wave function in the 3D space. 
These authors find for each particle an initial spin that is strictly zero and a variation of the spin module during the experiment from 0 to $\dfrac{\hbar}{2}$. This solution gives mathematically a causal interpretation of the EPR-B experiment, but variability of spin module causes it to lose its physical sense. 

With our assumptions, we consider two initial spin vectors $\mathbf{s}^A$ and  $\mathbf{s}^B$ with a module $\dfrac{\hbar}{2}$ as in the one-body case. It is the total spin of the singlet that is equal to zero. We therefore assume that, at the initial time, we know the wave functions (\ref{eq:A}) and (\ref{eq:B}) of the particles A and B. In the de Broglie-Bohm interpretation, we assume also that the intial position of particle A is
known ($x_0^A$, $y_0^A=0$, $z_0^A$) as well as of the particle B ($x_0^B$, $y_0^B=0$,  $z_0^B$).

It remains to determine the evolution of these wave functions and the trajectories of
particles A and B. 

Let's start with particle A.
Equation (\ref{eq:7densiteapreschampmagnetiqueAB}) shows that the density
of A is independent of that of B: it is equal to the density of a family
of free particles in a Stern-Gerlach apparatus, whose
initial spin orientation has been randomly chosen (it is exactly the density given by equation (\ref{eq:densiteapreschampmagnetique}) in the Appendix).
Since the particle A can be described by the initial wave function (\ref{eq:A}), we can assume
that its evolution is that to a free particle in a Stern-Gerlach apparatus, i.e.:
\begin{eqnarray}
\Psi^A(\textbf{r}_A, t_0+ \triangle t+ t )=
\cos\frac{\theta_0^A}{2} f^{+}(\textbf{r}_A,t)|+_{A}\rangle
+ \sin\frac{\theta_0^A}{2}e^{i
\varphi_0^A}f^{-}(\textbf{r}_A,t)|-_{A}\rangle\label{eq:fonctiondondeA}
\end{eqnarray}

For an initial polarization
($\theta_0^A,\varphi_0^A$) and an initial position ($z_0^A$), we
obtain, in the de Broglie-Bohm interpretation~\cite{Bohm1993}, an evolution of the position
($z_A(t)$) and of the spin orientation of A
($\theta^A(z_A(t),t)$)~\cite{Gondran2005b}. In the interval $[t_0,t_0 + \Delta t]$ during passage through the electromagnet, we obtain:
\begin{eqnarray}
 &&\frac{d z_A}{d t}=\frac{\mu_0 B'_{0} t}{ m} cos\theta(z_A,t)
 \nonumber\\
 &\text{with }&
 \tan \frac{\theta(z_A(t),t)}{2}=\tan\frac{\theta_0}{2} e^{- \frac{\mu_0 B'_{0} t^{2}z_A}
{2 m \sigma_0^{2}}}\label{eq:7trajectoiredanschampSetGA}
\end{eqnarray}
with the initial condition $z_A(t_0)=z^A_0$;  and in the interval
[$t_0 + \Delta t $, $t_0 + \Delta t +t$] ($t\geq0$) after passing through the electromagnet:
\begin{eqnarray}
 &&\frac{d z_A}{d t}=u \frac{\tanh(\frac{(z_\Delta + ut) z_A}{\sigma_{0}^{2}})+\cos \theta_0}{1+\tanh(\frac{(z_\Delta + ut) z_A}{\sigma_{0}^{2}})\cos
 \theta_0}
 \nonumber\\
 &\text{and }&\tan \frac{\theta(z_A(t),t)}{2}=\tan\frac{\theta_0}{2} e^{- \frac{(z_\Delta + ut)z_A}{\sigma_0^{2}}}.\label{eq:7trajectoireapreschampSetGA2}
\end{eqnarray}

It is this evolution of the polarization which is shown in Figure~\ref{fig:SetG-10traj} for the initial polarization ($\theta_0$=$\dfrac{\pi}{3}$).
The behavior of particle A is independent of B, whether the particle is entangled or not.

Let us now study  particle B. 
Equation (\ref{eq:7densiteapreschampmagnetiqueAB}) shows that the density
of B is independent of time and of the density of B: it is equal to the density of a family of free particles, which is constant in $x$ and $z$. Therefore we can assume that the particle B is immobile in $x$ and $z$: $z_B(t)=z_0^B$ and $x_B(t)=x_0^B$. Moreover, B follows a rectilinear classical
trajectory in y with $y_B(t)= v_0t$. 

Equation (\ref{eq:7psiexperience1}) shows that spins of A and
B remain opposite throughout step 1. The spin of a particle A is oriented gradually following the
position of the particle. The
spin of particle B follows that of A, while remaining opposite. Therefore, we can assume that
the orientation of B spin is driven by the orientation of A spin, like an interaction-at-a-distance:
\begin{equation}\label{eq:angleB}
\theta^B(t)= \pi - \theta(z_A(t),t)~~~~\textnormal{and}~~~~ \varphi^B(t)=
\varphi(z_A(t),t)- \pi.   
\end{equation}

Since the particle B can be described by the initial wave function (\ref{eq:B}), we can 
then associate to the particle B the wave function:
\begin{eqnarray}
\Psi^B(\textbf{r}_B, t_0+ \triangle t+ t )=
f(\textbf{r}_B)
\left(
\cos\frac{\theta^B(t)}{2} |+_{B}\rangle +
\sin\frac{\theta^B(t)}{2}e^{i \varphi^B(t)}|-_{B}\rangle\right).\label{eq:fonctiondondeB}
\end{eqnarray}

This wave funtion is specific, because it depends upon initial
conditions of A (positions and spins). The orientation of B spin is driven by that of particle A \textit{through the
singlet wave function}. Thus, the singlet wave function is the
non-local hidden variable.

Finally, during the first step, the singlet spinor in configuration space (\ref{eq:7psiexperience1}) can be replaced by the two single-particle spinors given by equations (\ref{eq:fonctiondondeA}) and (\ref{eq:fonctiondondeB}).

After the "measurement" of A at time $t_1= t_0 + \Delta
t + t_D$, if the A measurement is $+$ (respectively $-$), i.e. $ \theta(z_A(t),t)=+ \dfrac{\pi}{2}$  (resp. $-\dfrac{\pi}{2}$), we can deduce from equations (\ref{eq:angleB}) and (\ref{eq:fonctiondondeB}) that:  
\begin{equation}\label{eq:angleBB}
 \Psi^B(\textbf{r}_B, t_1 )=
f(\textbf{r}_B)e^{i \varphi^B(t_1)}|-_{B}\rangle  ~~~(\textnormal{resp.}~ f(\textbf{r}_B)|+_{B}\rangle )
\end{equation}
We have also $x^B(t_1)= x_0^B$ and $x^B(t_1)= z_0^B$. After this first "measurement", the second step of the EPR-B experiment is exactly the case of a single particle in a Stern-Gerlach
magnet $\mathcal{B}$ which is at an angle $\delta$ (resp. $\pi-\delta$) in relation to
$\mathcal{A}$.

\begin{figure*}
\begin{center}
\includegraphics[width=0.8\linewidth]{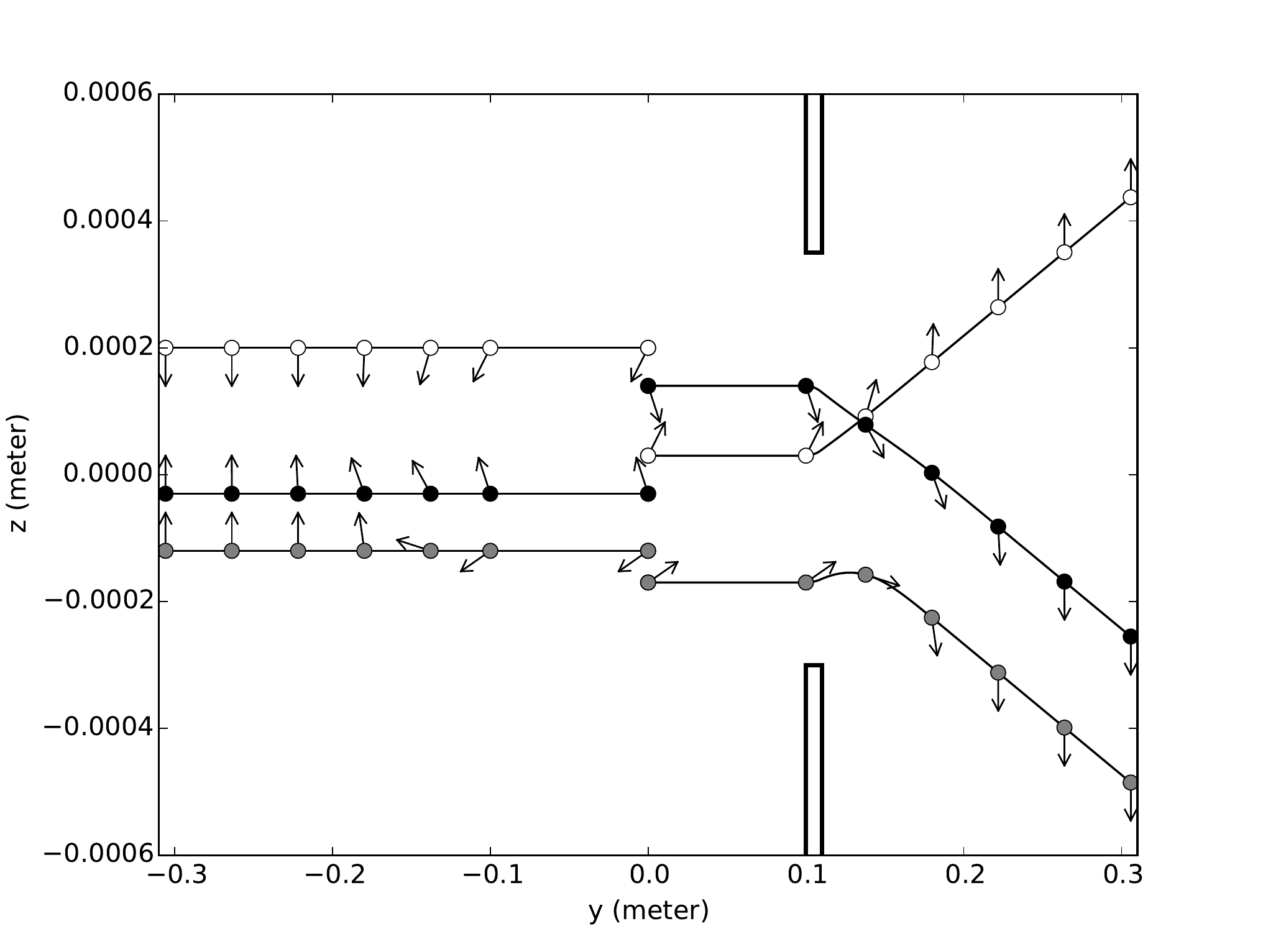}
\caption{\label{fig:EPR-5trajectoires}Evolution of the trajectories and the spin orientations (arrow) of three pairs of entangled A-B atoms (represented in black, gray and white) in the first step of the EPR-B experiment. There are created at the initial moment in $y=0$ ($y^A_0 =y^B_0=0$). $z^A_0$ and $z^B_0$
have been randomly chosen. $\theta^A_0$ is also randomly chosen, $\theta^B_0$ is in opposition to $\theta^A_0$, 
i.e.: $\theta^B_0= \pi - \theta^A_0$. Particle A goes to the right and cross a Stern-Gerlach device. Particle B goes to the left without crossing any device. The spin orientation along a trajectory ($\theta $) is indicated by an arrow.}
\end{center}
\end{figure*}

Figure~\ref{fig:EPR-5trajectoires} represents the evolution of trajectories and the spin orientations of three pairs of entangled A-B atoms in the first step of the EPR-B experiment. The three pairs of A-B particles are represented in black, gray and white on the figure. There are created at the initial moment in $y=0$ ($y^A_0 =y^B_0=0$).
The respective positions of A and B in relation to $z$, $z^A_0$ and $z^B_0$, at the initial moment have been randomly chosen. The orientation of the spin of A at the initial moment, $\theta^A_0$, is randomly chosen, in opposition to B, i.e.: $\theta^B_0= \pi - \theta^A_0$ (and $\varphi^B_0= \varphi^A_0 - \pi$). Particle A goes to the right and cross a Stern-Gerlach device. Particle B goes to the left without crossing any device. The spin orientation ($\theta $) is indicated by an arrow: the arrows on the right (respectively on the left) represent the evolution of the spin orientation of A (respectively of B).

\section{Conclusion}

We first recalled the definition of the spin vector in Bohmian mechanics (dependent on both  wave function and position) and its evolution during the phenomenon of measurement. In Bohmian mechanics, the "measured" value is not a preexisting value. It is the value obtained after a continuous orientation of the spin, either in the direction of the magnetic field gradient, or in the opposite direction.

Next, we have shown that, for the two entangled particles of the two-step version of the EPR-B experiment, it is possible to replace the singlet spinor in configuration space (\ref{eq:7psiexperience1}) by two single-particle spinors in physical space, given by equations (\ref{eq:fonctiondondeA}) and (\ref{eq:fonctiondondeB}).

We have demonstrated that the "first-measured" particle A behaves in a Stern-Gerlach apparatus as if it were not entangled. During the measurement of A, the particle density of B evolves as if it were not entangled. These two properties could be tested experimentally when the EPR-B experiment can be carried out easily.
This result was obtained in the de Broglie-Bohm interpretation using an integration of the two-body Pauli equation over time and space from an initial singlet with a spatial extension (\ref{eq:7psi-1}).

As de Broglie and Schr\"{o}dinger stated at the Solvay Conference in 1927, the wave function in configuration space may only be a mathematical tool that can be replaced by more physical wave functions. In our model, the A wave function is the same as that of a free particle in a Stern-Gerlach apparatus, the B wave function is the same as that of a free particle whose spin orientation vector is driven by the orientation of the A spin. Thus, we obtain a possible physical understanding of the EPR-B experience and the entanglement.

Our interpretation goes beyond the limit of the one put forward by Dewdney, Holland and Kyprianidis~ \cite{Dewdney1987b}, where EPR-B spins evolved from 0 to $\dfrac{\hbar}{2}$ from creation to measurement, which fits badly with spin quantification. We thus make the de Broglie-Bohm interpretation more credible.

These results also reopen the discussion about the completeness of quantum mechanics and the existence of hidden variables. Firstly, it clearly shows that Bohmian mechanics, which only uses resolution of the Pauli equation, gives the same statistical results as the Copenhagen interpretation for the Stern-Gerlach and EPR-B experiments.
It is the two-body Pauli equation that couples spin and spatial degrees of freedom in equations (\ref{eq:7psiexperience1}) and (\ref{eq:7psirotationBB}). Moreover, the measurement postulates and the postulate of wave packet reduction are not used in Bohmian mechanics and we show that they can be demonstrated (cf. Appendix).
 
The wave function of the singlet state alone introduces non locality: when we replace a singlet wave function in the configuration space with two wave functions in the 3D physical space, we must introduce \textit{interaction-at-a-distance} (equation (\ref{eq:angleB})) between the A spin orientation and that of B. 

\textit{Thus, the non-local influence in the EPR-B
experiment only concerns the spin orientation, not the motion
of the particles themselves.} This is a key point in the search for
a physical understanding of this non-local influence.

\section*{Appendix: Spin "measurement " in the Stern-Gerlach experiment}
 
\begin{figure}[H]
\begin{center}
\includegraphics[width=0.9\linewidth]{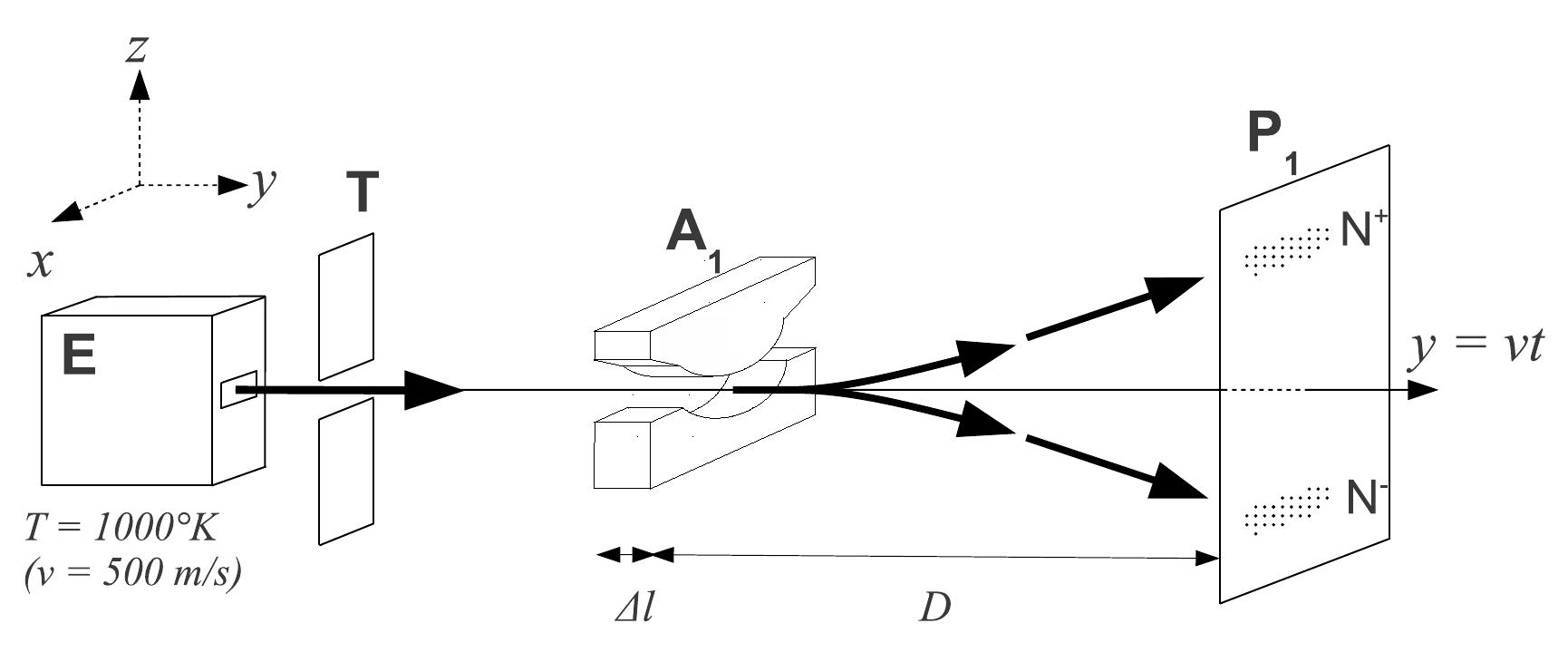}
\caption{\label{fig:schema-SetG} Schematic configuration of a Stern-
Gerlach apparatus.}
\end{center}
\end{figure}

The measurement of spin of a silver atom is carried out by a Stern-Gerlach  apparatus: an electromagnet $A$, where there is a strongly inhomogeneous magnetic field, followed by a screen $P$ (Fig.\ref{fig:schema-SetG}). In the Stern-Gerlach experiment, silver atoms contained in the oven E are heated to a high temperature and escape through a narrow opening. A second aperture, T, selects those atoms whose velocity, $v_0$, is parallel to the y-axis. The atomic beam passes through the gap of the electromagnet $A$, before  condensing on the screen $P$ on two spots of equal intensity $N^{+}$ and $N^{-}$. The magnetic moment of each silver atom before crossing the electromagnet is oriented randomly (isotropically). 
 
In the beam, we represent the atoms by their wave
function; one can assume that at the entrance to the
electromagnet $A$ (at the initial time $t=0$), each atom
can be approximatively described by a Gaussian spinor in $x$ and $z$:
\begin{equation}\label{eq:psi-0b}
    \Psi^{0}(x,z) = (2\pi\sigma_{0}^{2})^{-\frac{1}{2}}
                      e^{-\frac{x^2 +z^2}{4\sigma_0^2}}
                      \left( \begin{array}{c}\cos \frac{\theta_0}{2}e^{ i\frac{\varphi_0}{2}}
                                   \\
                                  \sin\frac{\theta_0}{2}e^{-i\frac{\varphi_0}{2}}
                  \end{array}
           \right)
\end{equation}
corresponding to a pure state. 
The variable y is treated in a classical way with $y=v_0 t$.

In (\ref{eq:psi-0b}), $\theta_0 $ and $\varphi_0 $ characterize the initial orientation of the spin. This initial orientation being randomized, on may suppose that $\theta_0 $ is drawn in a uniform law from [0, $\pi$] and that $\varphi_0 $ is drawn in a uniform law from [0, 2$\pi$]; In this way, we obtain a beam of atoms in which each atom has a different spinor: this is a model of a mixture of pure states.

In the Copenhagen interpretation, it is not necessary to resolve the Pauli equation. Just applying the postulates of quantum-mechanics measurement is sufficient. For measurement of the spin along the $z$-axis, the postulate of quantification states  that the measurement  corresponds to an eigenvalue of the spin operator $S_z =\frac{\hbar}{2}\sigma_z$, and the spectral decomposition postulate states that equation (\ref{eq:psi-s}) gives probability $cos^2 \frac{\theta}{2}$ (resp. $sin^2 \frac{\theta}{2}$) to measure the particle in the spin state $+ \frac{\hbar}{2}$ (resp.$ -\frac{\hbar}{2}$).

In the de Broglie-Bohm interpretation, the postulates of quantum mechanics measurement are not used, but demonstrated (see below). The results of the measurement are obtained, first by calculating the evolution of the wave function in interaction with the measuring apparatus with the Pauli equation (equation(\ref{eq:Pauli})), secondly by using the calculation of the wave function in space and time  to pilot the particle (equation (\ref{eq:vitesse})).

Let us consider the evolution of the initial wave function (\ref{eq:psi-0}) in the Stern-Gerlach apparatus. To obtain an explicit solution to the Stern-Gerlach experiment, we take the numerical values used in the Cohen-Tannoudji textbook~\cite{ CohenTannoudji1977}. For a
silver atom, we have $m = 1,8\times 10^{-25}$ kg, $v_0 = 500$\ m/s
, $\sigma_0$=10$^{-4}$m. For the electromagnetic field
$\textbf{B}$, $B_{x}=B'_{0} x$; $B_{y}=0$ and $B_{z}=B_{0} -B'_{0}
z$ with $B_{0}=5~Tesla$, $B'_{0}=\left| \frac{\partial B}{\partial
z}\right| =- \left| \frac{\partial B}{\partial x}\right|= 10^3~Tesla/m$ over a length $\Delta l=1~cm$. 

The variable $y$ will be treated classically with $y=v_0t$. The particle stays within the magnetic field for  a time $\Delta t= \frac{\Delta l}{v_0}=2 \times 10^{-5}s $. On exiting the magnetic field, the particle is free until it reaches screen $P$ placed at a $D=20~cm$ distance.

During this time
$[0,\Delta t]$, the spinor is  calculated (D\"{u}rr and al.\cite{Durr2004}, Gondran \cite{Gondran2005b}) with the Pauli equation (\ref{eq:Pauli}), where $\mu=\frac{e\hbar}{2m_e}$ is the Bohr magneton: 
\begin{equation}\label{eq:fonctiondanschampmagnetique}
\Psi (x,z,t)\simeq 
 \left(
\begin{array}{c}
                                \cos \frac{\theta_0}{2}
                 (2\pi\sigma_0^2)^{-\frac{1}{2}}
                 e^{-\frac{(z-\frac{\mu_{B} B'_{0}}{2 m}t^{2})^2 + x^2}
                 {4\sigma_0^2}} e^{i\frac{\mu_{B} B'_{0}t z -\frac{\mu^2_{0} B'^2_{0}}{6 m}t^3 +  \mu_B B_0 t + \frac{\hbar \varphi_0}{2}}{\hbar }}\\
                                i \sin \frac{\theta_0}{2}
                 (2\pi\sigma_0^2)^{-\frac{1}{2}}
                 e^{-\frac{(z+\frac{\mu_{B} B'_{0}}{2 m}t^{2})^2 + x^2}
                 {4\sigma_0^2}} e^{i\frac{-\mu_B B'_{0}t z -\frac{\mu^2_{0} B'^2_{0}}{6
    m}t^3 -  \mu_B B_0 t -\frac{\hbar \varphi_0}{2}}{\hbar }}
                            \end{array}
                     \right)
\end{equation}
After the magnetic field, at time $t+ \Delta t$ $(t \geq 0)$, in the free space, the
spinor becomes\cite{Gondran2005b}
\begin{equation}\label{eq:fonctionapreschampmagnetique}
\Psi (x,z,t+\Delta t)\simeq  \left(
\begin{array}{c}
                                \cos \frac{\theta_0}{2}(2\pi\sigma_0^2)^{-\frac{1}{2}}
                 e^{-\frac{(z-z_{\Delta}- ut)^2 + x^2}{2\sigma_0^2}} e^{i\frac{m u z + \hbar \varphi_+}{\hbar }} \\
                                i \sin \frac{\theta_0}{2} (2\pi\sigma_0^2)^{-\frac{1}{2}}
                     e^{-\frac{(z+z_{\Delta}+
                  ut)^2+ x^2}{2\sigma_0^2}}e^{i\frac{-
    muz + \hbar \varphi_-}{\hbar }}
                            \end{array}
                     \right)
\end{equation}

 where

\begin{equation}\label{eq:zdeltavitesse}
    z_{\Delta}=\frac{\mu_B B'_{0}(\Delta
    t)^{2}}{2 m}=10^{-5}m,~~~~~~u =\frac{\mu_B B'_{0}(\Delta t)}{m}=1 m/s.
\end{equation}

Equation (\ref{eq:fonctionapreschampmagnetique}) takes into
account the spatial extension of the spinor and we note that the
two-spinor components have very different values.

Since we have a mixture of pure states, the atomic density $\rho(z,t+\Delta t)$ is found by integrating $\rho(x, z,t+\Delta t) $  on $x$ and  on ($\theta_0$, $ \varphi_0$):
\begin{equation}\label{eq:densiteapreschampmagnetique}
    \rho(z,t+ \Delta t) =
     (2\pi\sigma_0^2)^{-\frac{1}{2}}
                  \frac{1}{2}\left(e^{-\frac{(z-z_{\Delta}- ut)^2}{2\sigma_0^2}}+
                  e^{-\frac{(z+z_{\Delta}+
                  ut)^2}{2\sigma_0^2}}\right).
\end{equation}

The decoherence time $t_D$, where the beam is separated into the two spots $N^{+}$ and $N^{-}$ ( when $z_{\Delta} +u t_D \geqslant 3\sigma_0$), is then given by the equation:
\begin{equation}\label{eq:tempsdecoherence}
t_{D} \simeq \frac{3 \sigma_{0}-z_\Delta}{u}=3 \times
10^{-4}s.
\end{equation}

We then obtain atoms with spins oriented only along the $z$-axis (positively
or negatively). 
Experimentally, we do not measure the spin directly, but
position ($\widetilde{x}$, $\widetilde{z}$) of the particle impact on $P$. If \ $\widetilde{z}\in N^+$, the term $\psi^-$ of
(\ref{eq:fonctionapreschampmagnetique}) is numerically equal to zero, and the
spinor $\Psi$ is proportional to $\binom{1}
                            {0} $, one of the eigenvectors of
$\sigma_z$~:
\begin{equation*}
\Psi (\tilde{z},t+\Delta t) \simeq (2\pi\sigma_0^2)^{-\frac{1}{4}}
\cos \frac{\theta_0}{2}
                 e^{-\frac{(\tilde{z}-z_{\Delta}- ut)^2 + \tilde{x}^2}
                 {4\sigma_0^2}} e^{i\frac{m u \tilde{z} + \hbar \varphi_+}{\hbar }}\left(
\begin{array}{c}
                                1 \\
                                0
                            \end{array}
                     \right).
\end{equation*}

If $\widetilde{z}\in N^-$, the term $\psi^+$ of
(\ref{eq:fonctionapreschampmagnetique}) is numerically equal to zero and the
spinor $\Psi $ is proportional to $\binom{0}
                            {1}$, the other eigenvector of $
\sigma_z $:
\begin{equation*}
\Psi (\tilde{z},t+\Delta t) \simeq
(2\pi\sigma_0^2)^{-\frac{1}{4}}\sin \frac{\theta_0}{2}
                e^{-\frac{(\tilde{z}+z_{\Delta}+ ut)^2+ \tilde{x}^2}
                {4\sigma_0^2}} e^{i\frac{-
    mu\tilde{z} + \hbar \varphi_-}{\hbar }}\left(
\begin{array}{c}
                                0 \\
                                1
                            \end{array}
                     \right).
\end{equation*}
Therefore, the measurement of the spin corresponds to an
eigenvalue of the spin operator $S_z= \frac{\hbar}{2} \sigma_z $.
It is a proof of the postulate of quantization in Bohmian mechanics.

Equation (\ref{eq:fonctionapreschampmagnetique}) gives the probability $ \cos^{2}
\frac{\theta_0}{2} $ (resp.$ \sin^{2} \frac{\theta_0}{2} $) of
measuring the particle in the spin state $+ \frac{\hbar}{2}$
(resp.$- \frac{\hbar}{2}$). It is a proof of the spatial decomposition
postulate in Bohmian mechanics.

\begin{figure}
\begin{center}
\includegraphics[width=0.7\linewidth]{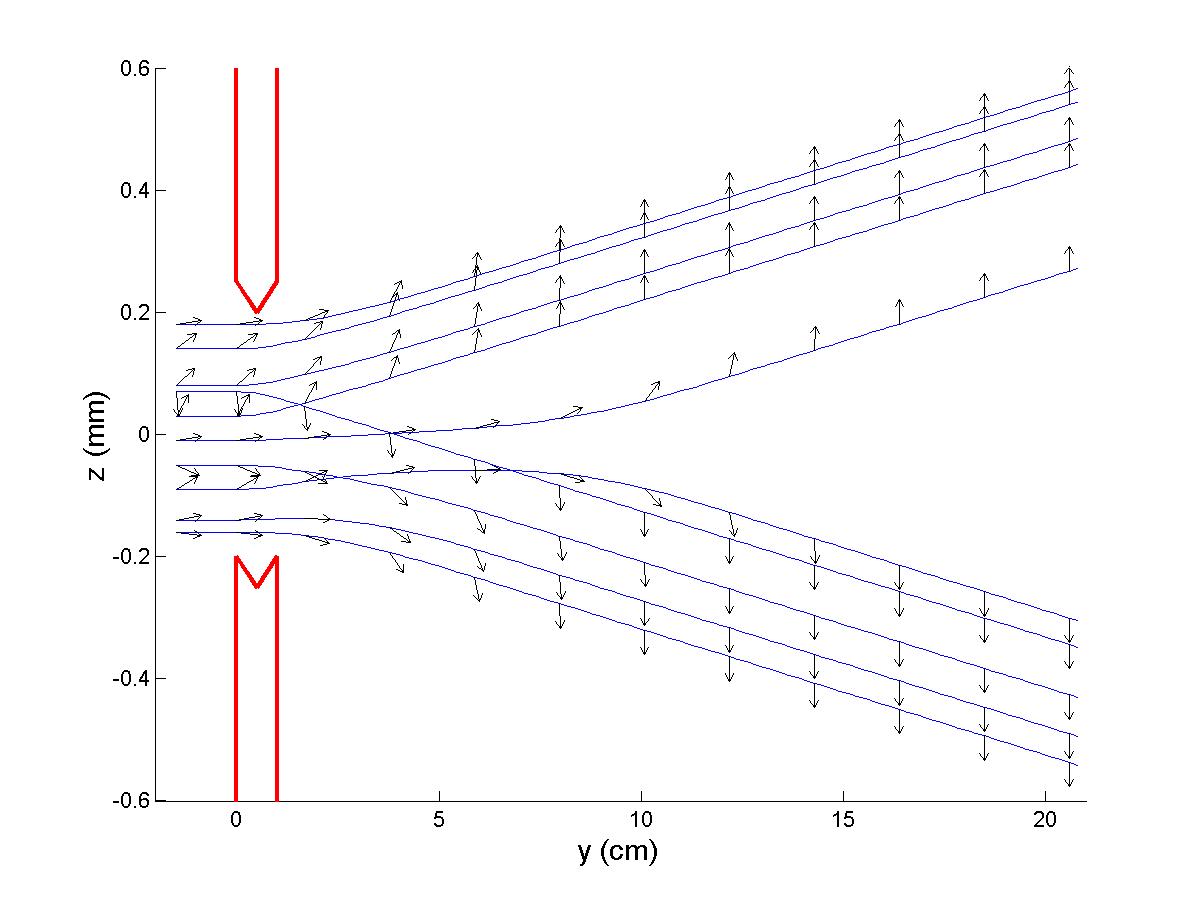}
\end{center}
\caption{\label{fig:EPRtraj} 10 silver atom
trajectories after the electro-magnet where the initial characteristics
$(\theta_0$,$\varphi_0$,$z_0$) have been randomly chosen; Arrows represent the
spin orientation $\theta(z(t),t)$.}
\end{figure}

Fig. \ref{fig:EPRtraj} presents in $x0y$ a set of 10
 silver-atom trajectories of which initial characteristics
$(\theta_0$,$\varphi_0$,$z_0$) have been randomly chosen:  $\theta_0$
and $\varphi_0$, which define on one hand the wave function, have uniform distributions, and $z_0^A$, which on the other hand defines the particle position in the wave function, has a normal distribution
$\mathcal{N}(0,\sigma_{0})$.
This double representation of quantum particles allows to take into account a mixture of pure states which satisfies the density of (\ref{eq:densiteapreschampmagnetique}).
 Spin orientation $\theta(z(t),t)$ is
represented by arrows.

We can see that the final orientation, obtained after the
decoherence time $t_{D}$, will depend on the initial particle
position $z_{0}^A$ in the wave packet and on the initial angle
$\theta_{0}^A$ of the atom magnetic moment with the $z$ axis.

Finally, we can also give a clear explanation of the Albert's example on contextuality~\cite{Albert1992} (p. 153-155). He considers a Stern-Gerlach experiment where the initial wave function is a pure state with symmetric spin orientation ($\theta_0 =\dfrac{\pi}{4}$ and so $z^{\theta_0}=0$). He changes, in a second experiment, the orientation of the magnetic field inside the Stern-Gerlach apparatus ($\textbf{B}$ into -$\textbf{B}$).
For the same initial position of atoms ( for example $z_0 > 0$), the first experiment gives a spin + and, in the second, a spin $-$. 
Bohmian mechanics explains this result: $u$ and $z_{\Delta}$ are proportional to $B'_0 =\dfrac{\partial B_{z}}{\partial z} $ (equation (\ref{eq:zdeltavitesse})), and therefore change their signs like $\textbf{B}$. By solving the Pauli equation, Bohmian mechanics is naturally contextual (involves the measuring device). 

\section*{Acknowledgement}
Authors gratefully acknowledge the two anonymous reviewers for their useful comments and constructive suggestions which helped to improve the paper.

\end{document}